\documentclass[aip,jap,amsmath,amssymb,twocolumn,10pt]{revtex4-1}
\usepackage{amsmath}
\usepackage{cases}
\usepackage{amssymb}
\usepackage{epstopdf}
\usepackage{graphicx}
\usepackage{color}
\usepackage{multirow}

\begin{document}
\title{Investigation of interface spacing, stability, band offsets and electronic properties on (001) $\rm SrHfO_3$/$\rm GaAs$ interface : First principles calculations}
\author{Li-Bin Shi}
\email{slb0813@126.com; shilibin@bhu.edu}
\author{Xiao-Ming Xiu}
\author{Xu-Yang Liu}
\author{Kai-Cheng Zhang}
\author{Chun-Ran Li}
\author{Hai-Kuan Dong}
\email{bhk@bhu.edu.cn}
\affiliation
{School of Mathematics and Physics, Bohai University, Liaoning Jinzhou 121013, China}
\date{\today}
\begin{abstract}
$\rm SrHfO_3$ is a potential dielectric material for metal-oxide-semiconductor (MOS) devices. $\rm SrHfO_3$/GaAs interface has attracted attention due to its unique properties. In this paper, the interface properties of (001) $\rm SrHfO_3$/GaAs are investigated by first principles calculations based on density functional theory (DFT). First of all, the adsorption behavior of Sr, Hf and O on GaAs surface is investigated. O has lower adsorption energy on Ga surface than on As surface. Then, some possible (0 0 1) $\rm SrHfO_3$/GaAs configurations are considered to analyze the interface spacing, stability, band offsets and charge transfer. $\rm HfO_2$/Ga(2) and $\rm SrO$/Ga(1) configurations in binding energy are lower than other interface configurations, indicating that they are more stable. At last, we study the electronic properties of $\rm HfO_2/Ga(2)$ and $\rm SrO/Ga(1)$ configurations. The electronic density of states suggests that the systems exhibit metallic behavior. The band offset and charge transfer are related to the interface spacing. The valence band offset (VBO) and charge transfer will decrease with increasing interface spacing.
\end{abstract}
\maketitle
\section{Introduction}
In order to reduce the feature size of metal-oxide-semiconductor-field-effect-transistors (MOSFETs), we have to replace traditional silicon dioxide gate dielectric with high permittivity (high-$\kappa$) materials.\cite{he2011integrations,Cho2010Structural} A lot of research has been done on high-$\kappa$ dielectrics, including $\rm Si_3N_4$,\cite{yang2009electronic} $\rm Ge_3N_4$,\cite{yang2008interface} $\rm Gd_2O_3$,\cite{hong1999epitaxial} $\rm La_2O_3$,\cite{chiu2005electrical} $\rm Y_2O_3$,\cite{wu2015single} $\rm Al_2O_3$,\cite{Choi2013Impact} $\rm ZrO_2$,\cite{Zheng2007First} $\rm HfO_2$,\cite{kang2003first-principles} $\rm SrTiO_3$,\cite{D2012first} and $\rm SrHfO_3$.\cite{liping2014first,liu2012first,liu2012structural,sawkar2010structural} Among them, $\rm SrHfO_3$ becomes a promising candidate because of its good current voltage characteristics, large band offsets, high dielectric constant, and low frequency dispersion. GaAs-based MOSFETs exhibit promising performance due to its higher electron velocity.\cite{he2013interface} It is found that single crystal interface usually exhibits excellent properties. For example, single crystal $\rm Gd_2O_3$/GaAs interface exhibits very low leakage current density with the value from $10^{-9}$ $\rm A/cm^2$ to $10^{-10}$ $\rm A/cm^2$. \cite{hong1999epitaxial} Cubic $\rm SrHfO_3$ has a experimental lattice constant of 4.07 \r{A}, which is approximately equal to 4.00 \r{A} in-plane spacing of GaAs.\cite{Mcdaniel2015Atomic}  This provides convenient conditions for the production of single crystal $\rm SrHfO_3$/$\rm GaAs$ interface.
\par
At present, some theoretical and experimental studies have been done on $\rm SrHfO_3$.\cite{Sousa2007Optical,Sousa2010Structural,Mcdaniel2015Atomic,rai2017electronic,Wang2016two} Mcdaniel et al.\cite{Mcdaniel2015Atomic} successfully deposited crystalline $\rm SrHfO_3$ films on Ge substrates. It is found that valence band offset (VBO) and conduction band offset (CBO) are 3.27 eV and 2.17 eV, respectively. Sousa et al.\cite{Sousa2007Optical} successfully deposited epitaxial $\rm SrHfO_3$ films on Si substrates. \textcolor[rgb]{0.00,0.07,1.00}{A band gap of 6.1 eV was measured optically. Sawkar-Mathur et al.\cite{Sousa2010Structural} successfully prepared $\rm SrHfO_3$ films on Si. The $\rm SrHfO_3$ exhibited a lattice mismatch of 6\% with silicon. Wang et al.\cite{Wang2016two} studied the electronic and structural properties of $\rm SrHfO_3$ by first principles calculations. The adsorption of Ga and As on $\rm SrO$ and $\rm HfO_2$ surfaces is systematically investigated. It is found that Ga and As preferentially adsorb at the top of O. In practice, the oxide is usually deposited on semiconductor substrate.\cite{konda2013high,lu2014band,Sousa2007Optical} It is more realistic to study the adsorption behaviors of Sr, Hf and O on GaAs surface. The adsorption behaviors of these atoms can provide useful information for determining interface stability. However, we still do not find a detailed theoretical or experimental study for $\rm SrHfO_3$ growth on GaAs.}
\par
In this study, we investigate the growth of $\rm SrHfO_3$ on GaAs. The adsorption behaviors of Sr, Hf and O on GaAs surface are studied. We deeply understand the stability of $\rm SrHfO_3$/GaAs by calculating binding energy. Band alignment between $\rm SrHfO_3$ and GaAs is investigated. The interface behaviors are analyzed by charge transfer and electronic density of states.
\section{Computational details}
The calculations on structural and electronic properties of $\rm SrHfO_3$ and GaAs are performed using the CASTEP code, which is based on density functional theory and plane-wave pseudopotential method. Local density approximation (LDA) is chosen as  exchange correlation functional. The valence electron configurations for Sr, Hf, Ga, As and O are considered as $4s^24p^65s^2$, $5d^26s^2$, $3d^{10}4s^24p^1$, $4s^24p^3$ and $2s^22p^4$, respectively. The cores are represented by norm-conserving pseudopotential, while the valences states are expanded in a plane-wave basis set with energy cutoff of 600 eV. The $6\times6\times6$ k point sampling is used for $\rm SrHfO_3$ and GaAs calculations, which is generated by the Monkhorst-Pack scheme.\cite{monkhorst1976special} \textcolor[rgb]{0.00,0.07,1.00}{The crystal structures of $\rm SrHfO_3$ and $\rm GaAs$ are shown in Fig. 1.} The geometry optimization will finish if the maximum force on each atom is less than 0.01 eV/\r{A}. \textcolor[rgb]{0.00,0.07,1.00}{The calculated lattice constant for $\rm SrHfO_3$ is 4.02 \r{A}, which is in agreement with theoretical value of 4.16\r{A} \cite{rai2017electronic} and experimental value of 4.07\r{A}. \cite{Mcdaniel2015Atomic} It is a little larger than in-plane spacing of GaAs with the value of 3.94 \r{A}.} The lattice mismatch between $\rm SrHfO_3$ and GaAs is estimated to be 2.0\%, which is agreement with the reported value of 1.8\%.\cite{Wang2016two} The band gaps calculated by LDA functional are $E_{\rm g}^{\rm GaAs(LDA)}$=0.78 eV and $E_{\rm g}^{\rm SrHfO_3(LDA)}$=3.60 eV. We attempt to improve the band gaps using hybrid Heyd-Scuseria-Ernzerhof functional (HSE).\cite{heyd2003hybrid,heyd2004efficient,paier2006erratum,heyd2005energy,Shi2017Firstprinciples}
The values calculated by HSE functional are $E_{\rm g}^{\rm GaAs(HSE)}$=1.58 eV and $E_{\rm g}^{\rm SrHfO_3(HSE)}$=5.37 eV, which are close to experimental values of  $E_{\rm g}^{\rm  GaAs}$=1.42 eV \cite{he2013interface,windhorn1982electron} and $E_{\rm g}^{\rm SrHfO_3}$ =6.01 eV.\cite{Mcdaniel2015Atomic}
\par
In order to investigate interface properties of (001) $\rm SrHfO_3$/$\rm GaAs$, we construct $2\times2$ surface supercell. Because $\rm SrHfO_3$ is usually grown on GaAs substrates, the lattice constants of the interface are determined by those of $\rm GaAs$. The interface consists of 11-layer $\rm SrHfO_3$ and 17-layer GaAs. A vacuum region of 15 \r{A} is used to avoid the interaction between the top and bottom layers. GaAs slab consists of alternating layers of Ga and As, while $\rm SrHfO_3$ slab consists of alternating layers of SrO and $\rm HfO_2$. In order to avoid polarization effect, $\rm SrHfO_3$ and GaAs are modeled using a symmetric slab. \cite{D2012first,Wang2016two} Ten kinds of interface configurations are considered to calculate, which is shown in Fig. 2. These configurations involve in depositing SrO or $\rm HfO_2$ layers on Ga or As surface. The atom located at the center of four Ga or As is called as fourfold hollow site. The atom located between two Ga or As is called as bridge site. For $\rm HfO_2$/Ga(1) interface, 4 Hf are located on the top of Ga, and 8 O at the bridge site. For $\rm HfO_2$/Ga(2) interface, 4 O are located on the top of the Ga, 4 O at the fourfold hollow site, and 4 Hf at the bridge site. For $\rm HfO_2$/Ga(3) interface, 8 O are located at the bridge site, and the 4 Hf at the fourfold hollow site. For $\rm SrO$/Ga(1) interface, 4 O are located on the top of Ga, and 4 Sr at the fourfold hollow site. For $\rm SrO$/Ga(2) interface, O and Sr are located at the bridge site. It is well known that the LDA functional underestimates the band gap of semiconductors and insulators. Therefore, the nonlocal functional is usually used to investigate the band gap.\cite{Shi2017Firstprinciples} However, the adsorption energy and binding energy only consider the system energy difference, so the result calculated by the LDA functional is reliable.\cite{Stankiewicz2016The,Ji2016Adsorption}
\par
The parameters for the interface calculation are different from those for the bulk calculation. $2\times2\times1$ k-points are used for integrations over the Brillouin zone. In the geometry relaxations, the atoms near interface and vacuum region are allowed to relax. The polarization inside $\rm SrHfO_3$ and GaAs slabs does not affect the atomic relaxation.

\section{Results and discussion}
\subsection{The (001) $\rm SrHfO_3$/GaAs interface}

\begin{figure}
\centering
\includegraphics[width=7 cm]{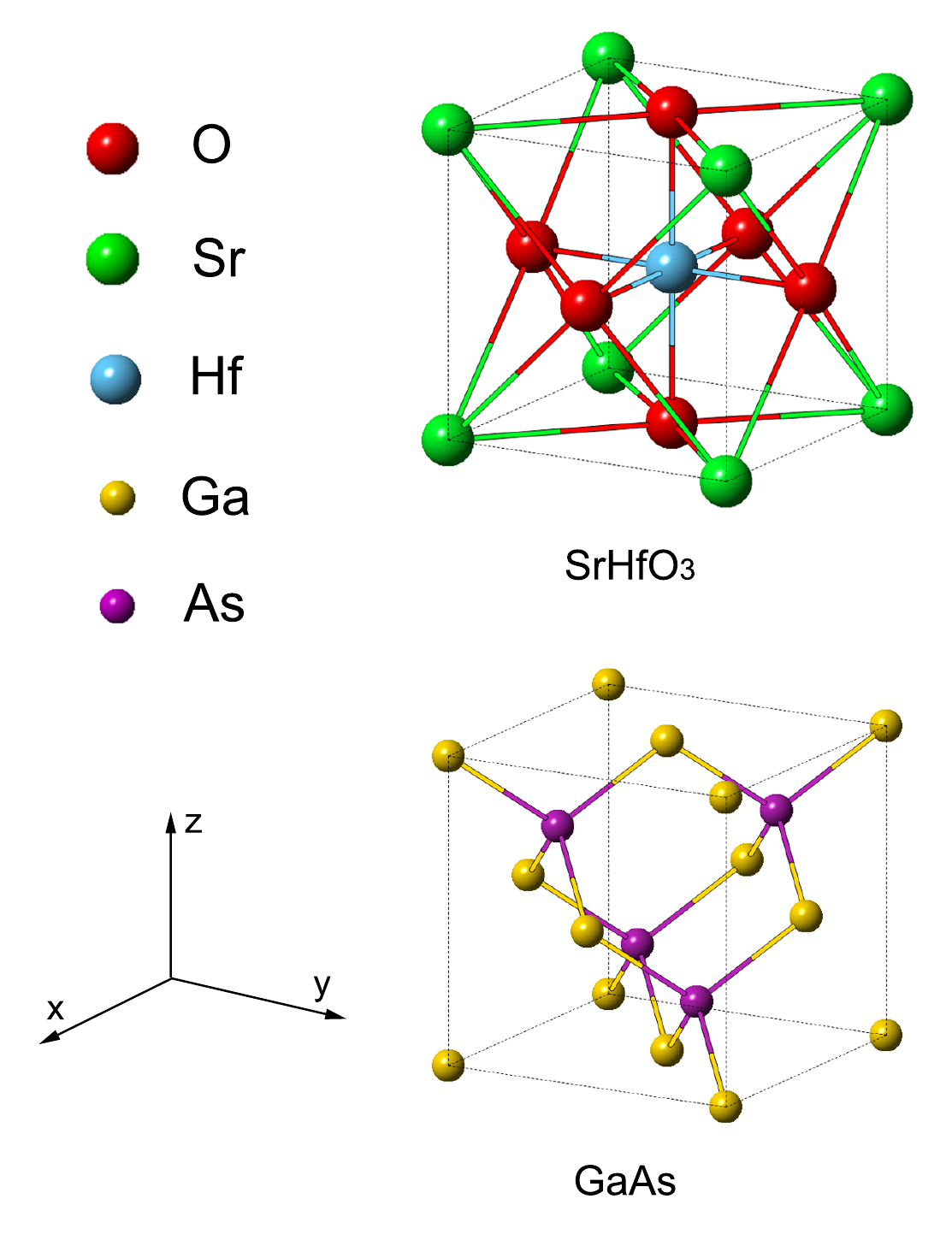}
\caption{Crystal structure of $\rm SrHfO_3$ and $\rm GaAs$. The coordinate axis is marked in the graph.}
\label{figure1}
\end{figure}

\begin{figure*}[htp]
\centering
\includegraphics[width=17cm]{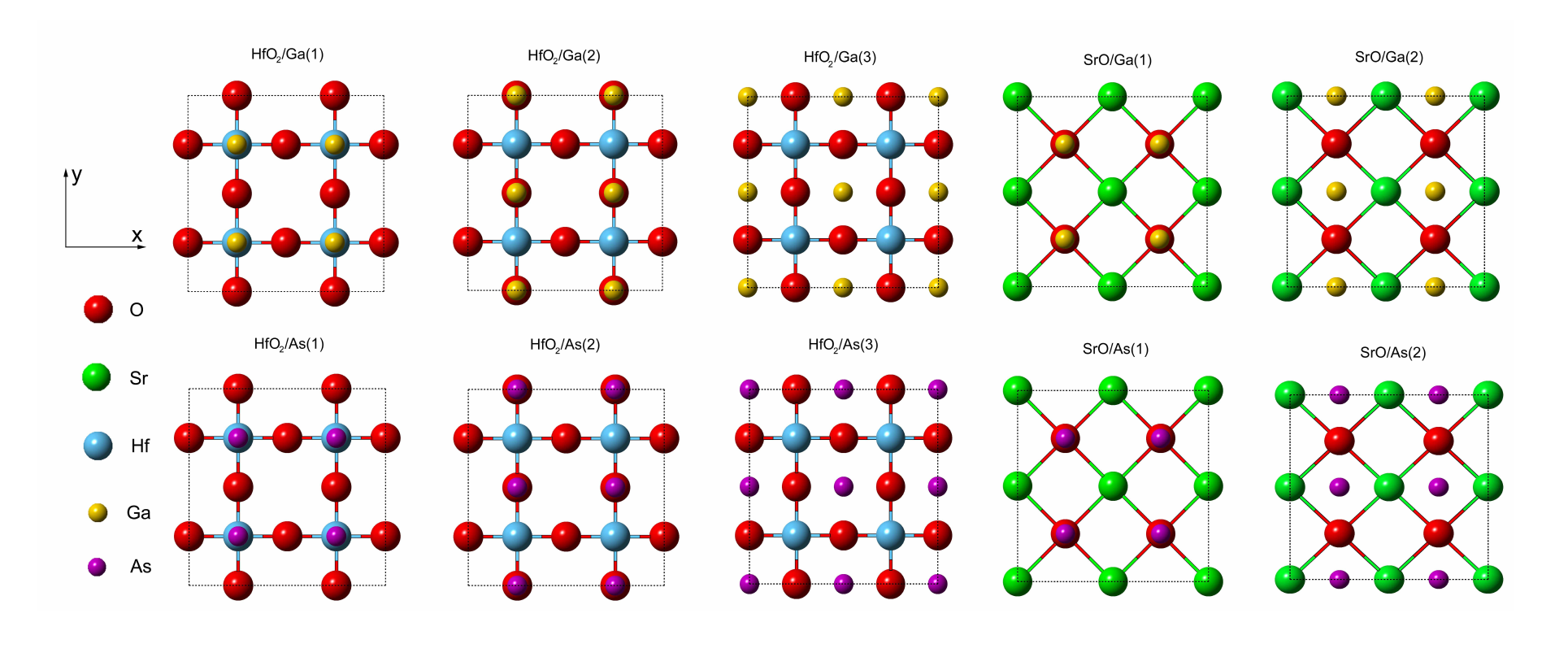}
\caption{The schematic of the (001) $\rm SrHfO_3$/GaAs interfaces. We assume that (001) $\rm SrHfO_3$ is deposited onto (001) GaAs substrate. 10 kinds of possible interface configurations are displayed. In order to facilitate the observation, we only give the atomic distribution at the interface.}
\label{figure2}
\end{figure*}

\begin{table}
\linespread{1.5}
\centering
  \caption{The adsorption energy per atom in units of electron volt(eV). T, B, and H denote top, bridge, and hollow sites, respectively.}
  \label{tb1}
  \footnotesize
  \begin{tabular}{cccc}
    \hline
    \hline
    Adsorption position              &$\rm O$ adatom                   &$\rm Sr$ adatom               &$\rm Hf$ adatom      \\
    \hline
    $T_{Ga}$                              &-7.24                             &                             &-4.00            \\
    $B_{Ga}$                              &-8.37                             &-3.29                        &-5.72            \\
    $H_{Ga}$                              &-7.34                             &-4.45                        &-9.21            \\
    $T_{As}$                              &-6.51                             &                             &-3.94            \\
    $B_{As}$                              &-7.17                             &-4.31                        &-6.72            \\
    $H_{As}$                              &-4.67                             &-5.46                        &-8.70            \\
    \hline
    \hline
\end{tabular}
\end{table}

\begin{table*}
\linespread{1.5}
\centering
  \caption{Optimized interface spacing, binding energy, band offsets (VBO and CBO), interface atom migration, and charge transfer. The value in parentheses is the interface spacing without atom relaxation.}
  \label{tb1}
  \footnotesize
  \begin{tabular}{c c c c c c c c c c c}
    \hline
    \hline
  Interface &Interface spacing (\r{A})  &Binding energy (eV/$\rm \r{A}^2$) &$\rm VBO$ (eV)&$\rm CBO$ (eV) &$\rm O $(\r{A}) &$\rm Hf $(\r{A})&$\rm Sr $(\r{A}) &$\rm Ga $(\r{A}) &$\rm As $(\r{A})  &$\rm Charge $ (e) \\
    \hline
    $\rm HfO_2/Ga(1)$      &2.8(2.9)     &-0.024    &2.89      &1.70      &0.44, -0.42   &-0.01         &       &0.07       &        &-0.70  \\
    $\rm HfO_2/Ga(2)$      &2.2(2.1)     &-0.038    &4.40      &0.19      &0.28, -0.47   &-0.07         &       &-0.06      &        &-1.08   \\
    $\rm HfO_2/Ga(3)$      &2.3(2.3)     &-0.017    &3.57      &1.02      &0.39, -0.47   &-0.03         &       &-0.07      &        &-0.72  \\
    $\rm SrO/Ga(1)$        &1.7(1.9)     &-0.048    &4.38      &0.21      &-0.02         &              &-0.14  &-0.19      &        &-0.40  \\
    $\rm SrO/Ga(2)$        &2.4(2.5)     &-0.017    &2.91      &1.68      &-0.02         &              &-0.203 &0.07       &        &-0.08  \\
    $\rm HfO_2/As(1)$      &2.9(2.9)     &-0.034    &2.50      &2.09      &-0.31         &0.04          &       &           &-0.07   &-0.24  \\
    $\rm HfO_2/As(2)$      &2.5(2.9)     &-0.017    &2.81      &1.78      &0.41          &-0.02         &       &           &-0.03   &-0.60  \\
    $\rm HfO_2/As(3)$      &2.8(2.8)     &-0.010    &3.04      &1.55      &0.38          &-0.05         &       &           &-0.08   &-0.32  \\
    $\rm SrO/As(1)$        &2.9(2.6)     &-0.008    &2.12      &2.76      &-0.03         &              &-0.18  &           &-0.02   &0.17  \\
    $\rm SrO/As(2)$        &2.8(2.8)     &-0.009    &1.92      &2.67      &-0.04         &              &-0.18  &           &-0.04   &0.20  \\
    \hline
    \hline
\end{tabular}
\end{table*}
Firstly, we study the adsorption behaviors of O, Sr and Hf on (001) GaAs slab. The (001) GaAs slab has a vacuum region of 15 \r{A}. The adsorption atom is gradually close to the absorption site. Adsorption energy is defined as the following expression:\cite{li2017first}
\begin{align}
\rm E_{ads}= E(GaAs+ads)-E(GaAs)- E(ads)
\end{align}
where E(GaAs+ads) is the total energy of a adsorption system with adatom, E(GaAs) is the total energy of the clean (001) GaAs slab, and E(ads) is the energy of adatom. We examine these adatoms at the fourfold hollow site, bridge site, and top site of Ga or As. Table 1 presents the calculated results. For Ga surface, Sr and Hf preferentially adsorb at Ga fourfold hollow site, while O preferentially adsorb at Ga bridge site. Similar results can also be found on As surface. It is noted that adsorption energy of O atom is lower on Ga surface than on As surface, which is in agreement with previous investigation.\cite{bakulin2016fluorine}\textcolor[rgb]{0.00,0.07,1.00}{ Previously, Wang et al.\cite{Wang2016two} investigated the adsorption behavior of Ga and As on SrO and $\rm HfO_2$ surface. They found that Ga and As preferentially adsorb at the top of O, and As is more favorable in energy than Ga. We investigate the growth behavior of $\rm SrHfO_3$ deposited on $\rm GaAs$ substrate, while they analyze the growth behavior of $\rm GaAs$ deposited on $\rm SrHfO_3$ substrate. This difference is mainly due to different film growth mechanisms. \cite{mallik2017effect,kacher2016initial}}
\par
In order to investigate the stability of (001) $\rm SrHfO_3$/GaAs interface, we also calculate the interface binding energy, which is defined as \cite{zhang2014first}
\begin{align}
\rm E_b(SrHfO_3/GaAs)=[&\rm E_i (SrHfO_3/GaAs)-E_{s}(SrHfO_3)\notag\\&-\rm E_{s}(GaAs)]/2S
\end{align}
where $\rm E_i$ ($\rm SrHfO_3$/GaAs) is total energy of $\rm SrHfO_3$/GaAs interface, $\rm E_s$($\rm SrHfO_3$) and $\rm E_s$(GaAs) denote total energy of $\rm SrHfO_3$ and GaAs surface, respectively, S is the cross-sectional area of the interface, and the factor 2 denotes two identical interfaces in the supercell. In this study, we accurately determine the interface spacing by two steps. We first roughly estimate the interface spacing by total energy calculation. We do not allow the atoms to relax, only calculate the total energy of the system at different interface spacing. The interface spacing is changed in the range from 1 \r{A} to 5 \r{A} with the interval of 0.1 \r{A}. Figure 3 presents binding energy versus interface spacing. Although the symmetrical slab is adopted in the calculation, the binding energy curves can still reflect the binding characteristics of different interfaces. \textcolor[rgb]{0.00,0.07,1.00}{The lowest binding energy and interface spacing are -0.022 eV/$\rm\r{A}^2$ and 2.9 \r{A} for $\rm HfO_2$/Ga(1), -0.031 eV/$\rm \r{A}^2$ and 2.1 \r{A} for $\rm HfO_2$/Ga(2), -0.015 eV/$\rm \r{A}^2$  and 2.3 \r{A} for $\rm HfO_2$/Ga(3), -0.049 eV/$\rm \r{A}^2$ and 1.9 \r{A} for $\rm SrO$/Ga(1), -0.019 eV/$\rm \r{A}^2$ and 2.5 \r{A} for $\rm SrO$/Ga(2), -0.019 eV/$\rm \r{A}^2$ and 2.9 \r{A} for $\rm HfO_2$/As(1), -0.011 eV/$\rm \r{A}^2$ and 2.9 \r{A} for $\rm HfO_2$/As(2), -0.008 eV/$\rm \r{A}^2$ and 2.8 \r{A} for $\rm HfO_2$/As(3), -0.012 eV/$\rm \r{A}^2$ and 2.6 \r{A} for $\rm SrO$/As(1), -0.012 eV/$\rm \r{A}^2$ and 2.8 \r{A} for $\rm SrO$/As(2). }The negative binding energy indicates that the interface is stable. It is found that the binding energies for $\rm HfO_2$/Ga(2) and $\rm SrO$/Ga(1) are lower than those for other interface configurations, which suggests that $\rm HfO_2$/Ga(2) and $\rm SrO$/Ga(1) interfaces are more stable. The binding energy curve of $\rm HfO_2$/As(2) interface exhibits abnormal peak, which is independent of the chosen exchange correlation functional and neglecting spin polarization. This abnormal behavior may be due to the competition between Hf-As and O-As interactions. The binding energy is -0.325 eV/$\rm \r{A}^2$ for CoO/MnO interface \cite{yao2008first}, -0.033 eV/$\rm \r{A}^2$ for $\rm Cu_2ZnSnS_4$/ZnO interface \cite{cheng2016bonding}, and -0.315 eV/$\rm \r{A}^2$ for Mg/$\rm Al_4C_3$ interface. \cite{li2013first}
\par
\begin{figure}[htp]
\centering
\includegraphics[width=8 cm]{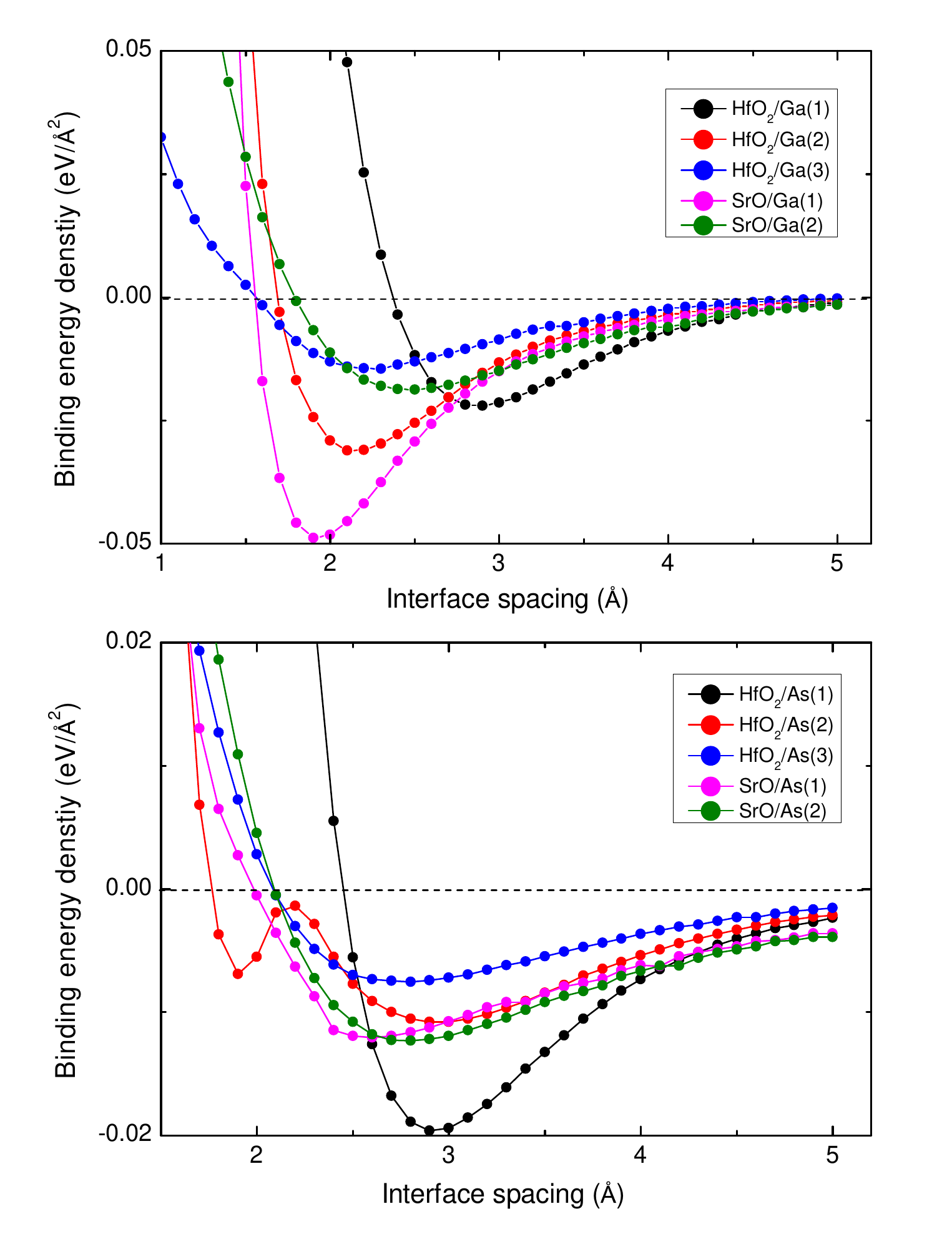}
\caption{Binding energy versus interface spacing for ten kinds of possible interface configurations.}
\label{figure3}
\end{figure}
\par
In order to accurately determine the interface spacing, we allow the atoms near interface and vacuum region to fully relax. The system has the lowest energy at the optimized interface spacing. Therefore, interface spacing is determined by the lowest energy. The value is 2.8 \r{A} for $\rm HfO_2$/Ga(1), 2.2 \r{A} for $\rm HfO_2$/Ga(2), 2.3 \r{A} for $\rm HfO_2$/Ga(3), 1.7 \r{A} for $\rm SrO$/Ga(1), 2.4 \r{A} for $\rm SrO$/Ga(2), 2.9 \r{A} for $\rm HfO_2$/As(1), 2.5 \r{A} for $\rm HfO_2$/As(2), 2.8 \r{A} for $\rm HfO_2$/As(3), 2.9 \r{A} for $\rm SrO$/As(1), and 2.8 \r{A} for $\rm SrO$/As(2). Table 2 gives the binding energy, band offsets, atom migration, and charge transfer under optimized interface spacing. The values in parentheses are the interface spacing without atom relaxation. Some of the interface spacings slightly change after atom relaxation. The interface spacing for stable heterojunction is normally in the range of 1.5 \r{A}$\sim$ 3.0 \r{A}.\cite{He2015First,Luo2015Electronic} The binding energy only slightly changes after the atom relaxation except for the $\rm HfO_2$/As(1) interface. The binding energies for $\rm HfO_2$/Ga(2) and $\rm SrO$/Ga(1) are still lower than those for other interface configurations.
\par
The atoms near interface will migrate after atom relaxation. The positive value indicates that the atom moves toward the interface, while the negative value indicates that the atom moves away from the interface. The results show that O has a larger migration on $\rm HfO_2$ layer than on $\rm SrO$ layer. Hf has smaller migration than Sr. The migration of Ga and As is less than 0.1 \r{A}. The charge in Table 2 represents the charge transfer from GaAs slab to $\rm SrHfO_3$ slab, which is the number of extra electrons on $\rm SrHfO_3$ slab. The interface characteristics can be investigated by the analysis of charge transfer.\cite{Stefan2016Interface} It is noted that $\rm HfO_2$ layer is easier to get electrons than $\rm SrO$ layer. This may be due to the fact that $\rm HfO_2$ layer has more oxygen atoms than $\rm SrO$ layer. The charge is -0.70 e for $\rm HfO_2$/Ga(1), -1.08 e for $\rm HfO_2$/Ga(2), -0.72 e for $\rm HfO_2$/Ga(3), -0.40 e for $\rm SrO$/Ga(1), and -0.08 e for $\rm SrO$/Ga(2). Because electronegativity of As is greater than that of Ga, the charge transfers at $\rm HfO_2$/As and $\rm SrO$/As interfaces are smaller than those at $\rm HfO_2$/Ga and $\rm SrO$/Ga interfaces. Anomalous charge transfer is observed for $\rm SrO$/As(1) and $\rm SrO$/As(2) interfaces. The charge is 0.17 e for $\rm SrO$/As(1), 0.20 e for $\rm SrO$/As(2), which suggests that As obtains electrons from SrO layer.
\par
In the potential alignment method, the valence band offset (VBO) is determined by two terms.\cite{silvestri2013first,D2012first}
\begin{equation}
\rm VBO=\Delta E_v+\Delta V
\end{equation}

The first contribution of $\rm \Delta E_v$ corresponds to alignment of valence band maximum (VBM) for bulk band structure term. The second term of $\rm \Delta V$ corresponds to the macroscopic averaged electrostatic potential (AEP) alignment. Because of the error between the calculated and experimental band gap, we can't get the exact CBO by calculated band gap. In this study, we obtain CBO by the relation CBO=VBO+$E_{\rm g}^{\rm SrHfO_3(EXP)}$-$E_{\rm g}^{\rm GaAs(EXP)}$.\cite{cheng2016bonding} Table 2 shows that the VBO and CBO are closely related to the interface configurations, which is in agreement with previous investigation.\cite{zhang2003atomic,hajlaoui2015first,dong2005first,bao2013band} The VBO and CBO are determined to be 2.89 eV and 1.70 eV for $\rm HfO_2$/Ga(1), 4.40 eV and 0.19 eV for $\rm HfO_2$/Ga(2), 3.57 eV and 1.02 eV for $\rm HfO_2$/Ga(3), 4.38 eV and 0.21 eV for $\rm SrO$/Ga(1), 2.91 eV and 1.68 eV for $\rm SrO$/Ga(2), 2.50 eV and 2.09 eV for $\rm HfO_2$/As(1), 2.81 eV and 1.78 eV for $\rm HfO_2$/As(2), 3.04 eV and 1.55 eV for $\rm HfO_2$/As(3), 2.12 eV and 2.76 eV for $\rm SrO$/As(1), 1.92 eV and 2.67 eV for $\rm SrO$/As(2). \textcolor[rgb]{0.00,0.07,1.00}{The VBO and CBO for ideal MOS devices should be larger than 1 eV, which can effectively prevent leakage current. Previous investigation showed that the VBO and CBO are 2.66 eV and 1.33 eV for $\rm ZrO_2$/GaAs. \cite{WANG2015Impact} Yang et al. found that VBO for $\rm La_2O_3$/GaAs interface is about 1.7 eV. \cite{yang2005energy} Jin et al. found that the experimental VBO changes from 2.86 eV to 3.36 eV when the annealing temperature increases from  600 $^{\circ}$C to 800 $^{\circ}$C. \cite{Jin2005Temperature}}
\par
\textcolor[rgb]{0.00,0.07,1.00}{In this study, we do not consider the effect of interface defects. Defects inevitably exist in the actual interface. Previously, Alay-e-Abbas et al. investigated formation energies of vacancy defect in $\rm SrHfO_3$. \cite{Alayeabbas2014Thermodynamic} The results show that ordered oxygen vacancies in $\rm HfO_2$ layer are energetically favorable. Chen et al did a systematic study of stability diagram and electronic properties for (110) $\rm SrHfO_3$ polar terminations. \cite{Chen2016First} They found that the atomic defect on the surface layer dominates the stabilization for $\rm SrHfO_3$. We believe that the defects near $\rm SrHfO_3$/$\rm GaAs$ interface will affect the band offset, which should be further studied.}

\begin{figure*}
\centering
\includegraphics[width=14cm]{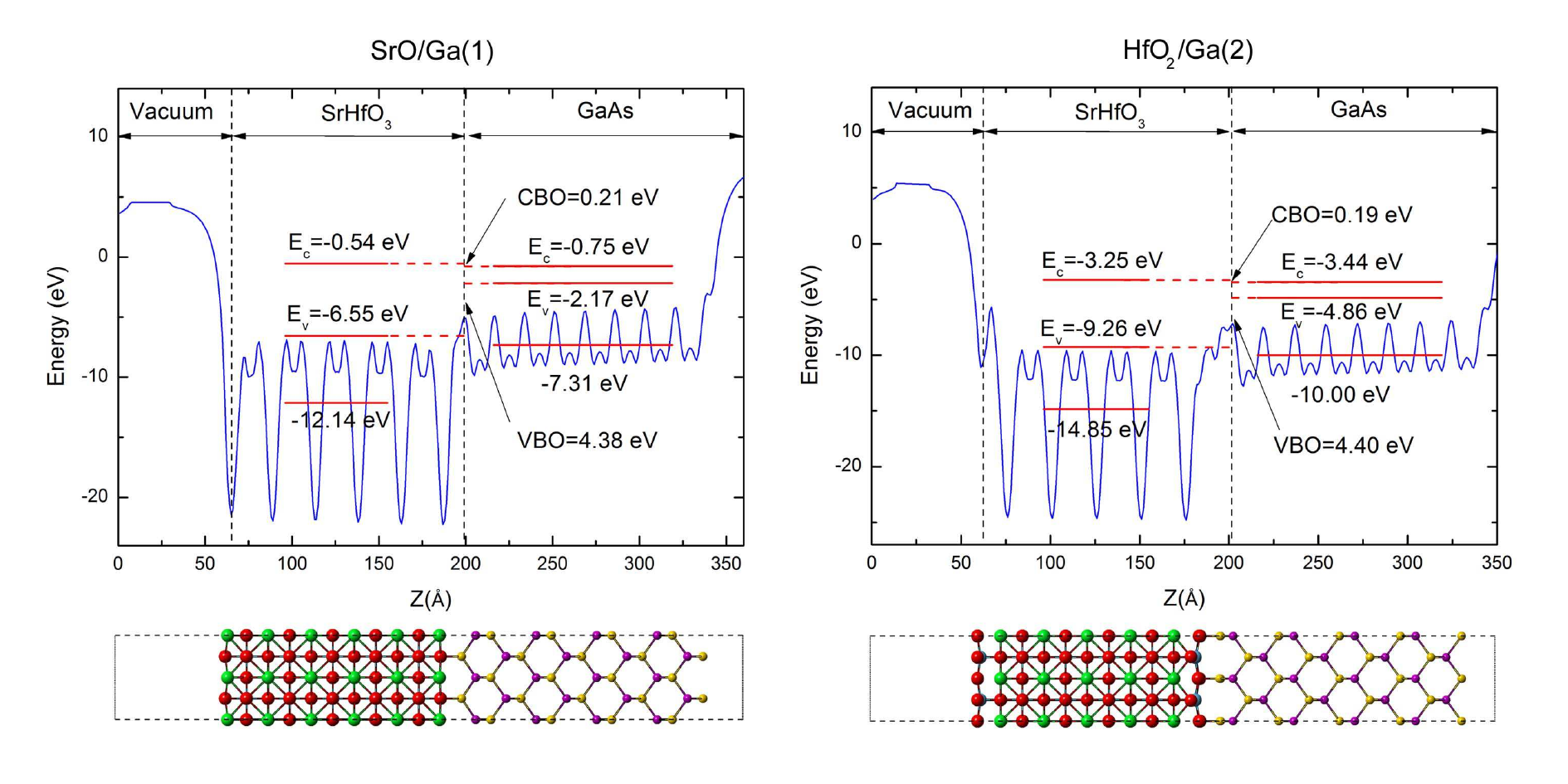}
\caption{Schematic representation of the band alignment for $\rm SrO/Ga(1)$ and $\rm HfO_2/Ga(2)$ configurations (top panel). Blue curve denotes planar averaged electrostatic potential (AEP). Macroscopic AEP, valence band maximum ($E_v$), conduction band minimum ($E_c$), band offsets (VBO and CBO) are also presented in figure.
A schematic view of the $\rm SrHfO_3$/GaAs interface is shown in the bottom panel. The Sr, Hf, O, Ga and As atoms are marked by large green, large blue, large red, small yellow and small violet spheres,respectively.}
\label{figure4}
\end{figure*}

\subsection{Interface properties of $\rm HfO_2/Ga(2)$ and $\rm SrO/Ga(1)$ configurations}

Figure 4 presents a schematic representation of the band alignment for $\rm HfO_2/Ga(2)$ and $\rm SrO/Ga(1)$. The planar and macroscopic AEP are presented in Fig. 4. The planar AEP in the atomic region shows periodic oscillations, while it keeps constant in the vacuum region (see blue curve), which is in agreement with previous investigation. \cite{yang2008interface,yang2009electronic,Yang2016Interfacial} In order to accurately calculate band offsets, we have to eliminate the interface polarization effect. The polarization inside $\rm SrHfO_3$ and $\rm GaAs$ can be avoided by a symmetric slab. During calculations, the dipole correction is also applied to remove the possible interface polarization induced by charge transfer between the oxide and semiconductor. For $\rm HfO_2/Ga(2)$, macroscopic AEP value is -12.14 eV for $\rm SrHfO_3$ and -7.31 eV for GaAs. $\rm E_v$ (valence band maximum) and $\rm E_c$ (conduction band minimum) are -6.55 eV and -0.54 eV for $\rm SrHfO_3$ and -2.17 eV and -0.75 eV for GaAs.  For $\rm SrO/Ga(1)$ interface, the value of macroscopic AEP is -14.85 eV for $\rm SrHfO_3$ and -10.00 eV for GaAs. $\rm E_v$ and $\rm E_c$ are -9.26 eV and -3.25 eV for $\rm SrHfO_3$ and -4.86 eV and -3.44 eV for GaAs. The VBO and CBO for $\rm HfO_2/Ga(2)$ and $\rm SrO/Ga(1)$ are also shown in Fig. 4.
\par
We investigate the VBO and charge transfer versus interface spacing for $\rm HfO_2$/Ga(2) and $\rm SrO$/Ga(1) interfaces, which is shown in Fig. 5. For $\rm HfO_2$/Ga(2) interface, VBO keeps small change as interface spacing is in the range from 2.1 \r{A} to 2.8 \r{A}. This reflects strong interaction between $\rm SrHfO_3$ and GaAs. VBO decreases as the interface spacing is in the range from 2.8 \r{A} to 3.2 \r{A}, which corresponds to the weakening interface interaction. The result can also be verified by the charge transfer. It is found that the charge on $\rm SrHfO_3$ slab decreases with the increase of the interface spacing. The charge changes from -1.08 e to -0.60 e as the interface spacing increases from 2.1 \r{A} to 3.2 \r{A}. Similar behavior is also found on $\rm SrO$/Ga(1). Comparing with $\rm HfO_2$/Ga(2), the charge transfer at $\rm SrO$/Ga(1) is small. It is noted that the interface spacing affects the band offsets, resulting in deviation between theoretical calculation and experiment.\cite{Tao2014The,Shi2017First,Shi2016Investigation,Li2017The}
\par
The layer electronic density of states for $\rm SrO/Ga(1)$ and $\rm HfO_2/Ga(2)$ is presented in Fig.6. The chemical formula of symmetric slab is $\rm Hf_{20}Sr_{24}O_{64}/Ga_{36}As_{32}$ for SrO/Ga(1) and $\rm Hf_{24}Sr_{20}O_{68}/Ga_{36}As_{32}$ for $\rm HfO_2/Ga(2)$. In order to elucidate the effect of the polarization on the interfacial properties, we also investigate $\rm SrHfO_3$/GaAs interface with ideal stoichiometry. The chemical formula is $\rm Hf_{20}Sr_{20}O_{60}/Ga_{32}As_{32}$ for SrO/Ga(1) and $\rm HfO_2/Ga(2)$ configurations, which is also called as unsymmetric slab. It is found that $\rm SrO/Ga(1)$ and $\rm HfO_2/Ga(2)$ exhibit metallic behavior, which is in agreement with previous investigation.\cite{Wang2016two} The metallic behavior for $\rm HfO_2/Ga(2)$ and $\rm SrO/Ga(1)$ occurs only on several atom layers near the interface. The polarization inside $\rm SrHfO_3$ and GaAs does not affect the interface behavior. The gap states in $\rm SrHfO_3$ are mainly determined by Hf 5d, Sr 4d, O 2p. The gap states in GaAs is dominated by Ga 4s, 4p and As 4s, 4p. \textcolor[rgb]{0.00,0.07,1.00}{We believe that the metal interface is mainly caused by the interfacial charge transfer, which will lead to carrier tunneling and increasing the leakage current of the electronic devices. \cite{Chang2015Impact,yeo2000direct} Previously, wang et al. explained that the metallicity in $\rm SrHfO_3$/GaAs interface originates from the surface electron accumulation. \cite{Wang2016two} Alay-e-Abbas et al found that the metal behavior in $\rm SrHfO_3$ is due to charge transfer between vacancy site and the hafnium dangling bond. \cite{Alayeabbas2014Thermodynamic} In order to eliminate the metal interface, the electrons near the interface should form stable covalent bonds. The interface issues still need to be further studied.}
\par
\textcolor[rgb]{0.00,0.07,1.00}{To further understand the bonding, the electron density difference of $\rm SrO/Ga(1)$ is given in Fig. 7. The charge density is mainly distributed around O, while Sr and Hf have almost no charge density. In GaAs slab, the charge density is mainly distributed between Ga and As. Because As has a greater electronegativity than Ga, the charge density is slightly closer to As. This result strongly suggests that the ionic bond is formed between O and Sr (Hf), while polar covalent bond is formed between Ga and As. We also test the effect of polar $\rm SrHfO_3$ and GaAs slabs on electron density. We can find no significant change in electron density using polar $\rm SrHfO_3$ and GaAs slabs. Similar behavior is also found at $\rm HfO_2/Ga(2)$ interface. This analysis is consistent with previous investigations.\cite{Sun2016First,Liu2017A,Xian2016Interfacial} The charge transfers for $\rm SrO/Ga(1)$ and $\rm HfO_2/Ga(2)$ interfaces occur mainly between interfacial O and Ga. O near interface will obtain more electrons than those in bulk $\rm SrHfO_3$, while Ga near interface will lose more electrons than those in bulk $\rm GaAs$. The interaction between interfacial O and Ga should be a typical ionic bond.}

\section{Conclusions}
Geometry structures and electronic properties of $\rm SrHfO_3$/GaAs interface are investigated by density functional theory (DFT). O is more favorable for adsorption on Ga surface than on As surface. Ten possible interface configurations on $\rm SrHfO_3$/GaAs interface are investigated. The results show that the band offsets, interface spacing, binding energy and charge transfer are different for ten configurations. The interface spacing is in the range from 1.7 \r{A} to 2.9 \r{A}, and the charge transfer is in the range from 0.20 e to -1.08 e. Both $\rm HfO_2/Ga(2)$ and $\rm SrO/Ga(1)$ interfaces are lower in binding energy than other interface configurations, which suggests that they have a stable structure. Electronic density of states for $\rm HfO_2/Ga(2)$ and $\rm SrO/Ga(1)$ suggests that they exhibit metallic interface behavior. The charge transfer occurs mainly from interfacial Ga to O, which suggests that the interaction between interfacial O and Ga is typical ionic bond. As the interface spacing increases, VBO begins to undergo small changes and then monotonically decreases. We also find that the charge transfer decreases with the increase of interface spacing, which corresponds to the weakening interface interaction.

\begin{figure}[htp]
\centering
\includegraphics[width=8.5 cm]{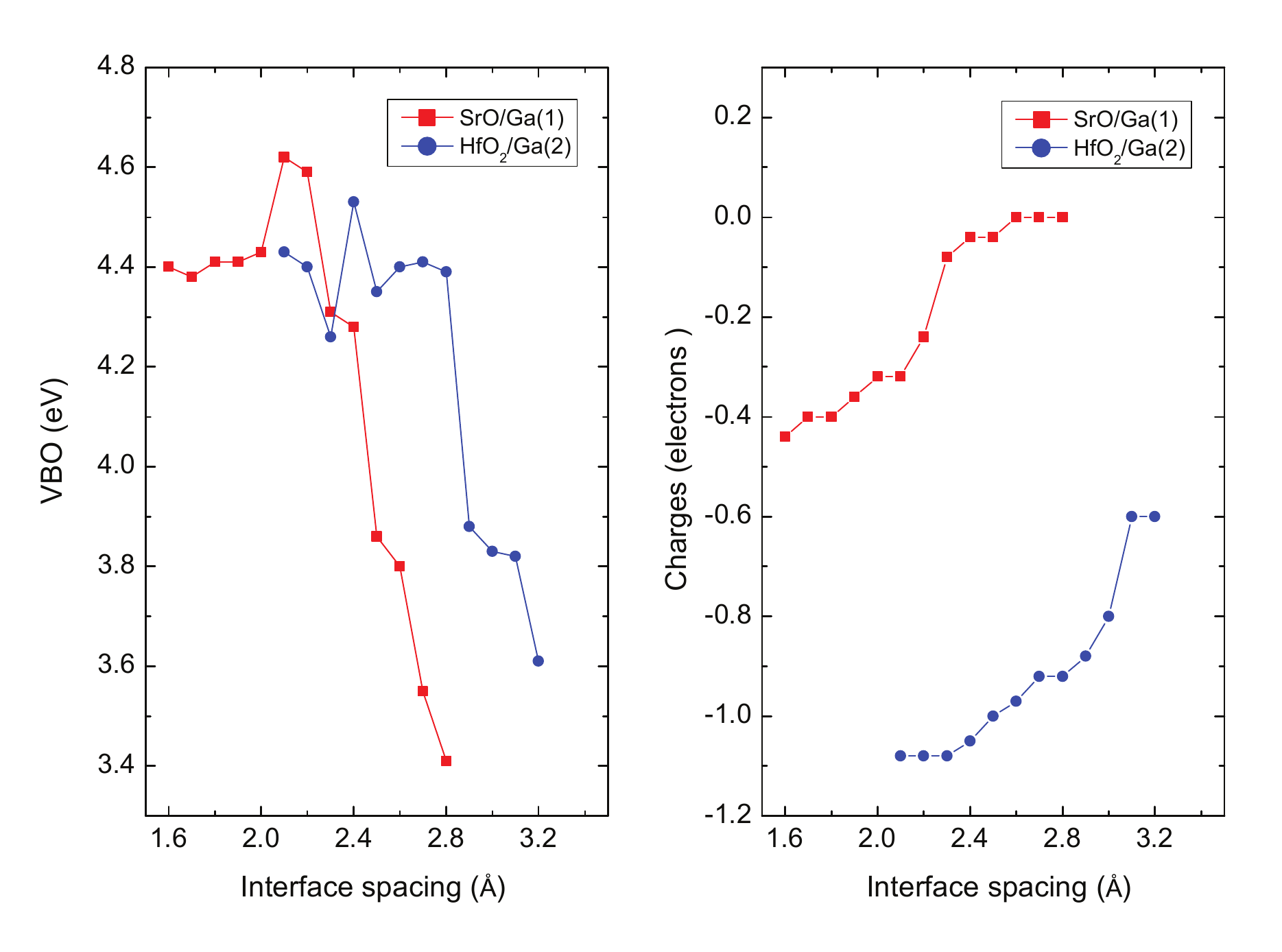}
\caption{VBO and charges versus interface spacing for $\rm SrO/Ga(1)$ and $\rm HfO_2/Ga(2)$ configurations.}
\label{figure5}
\end{figure}

\begin{figure*}
\centering
\includegraphics[width=14cm]{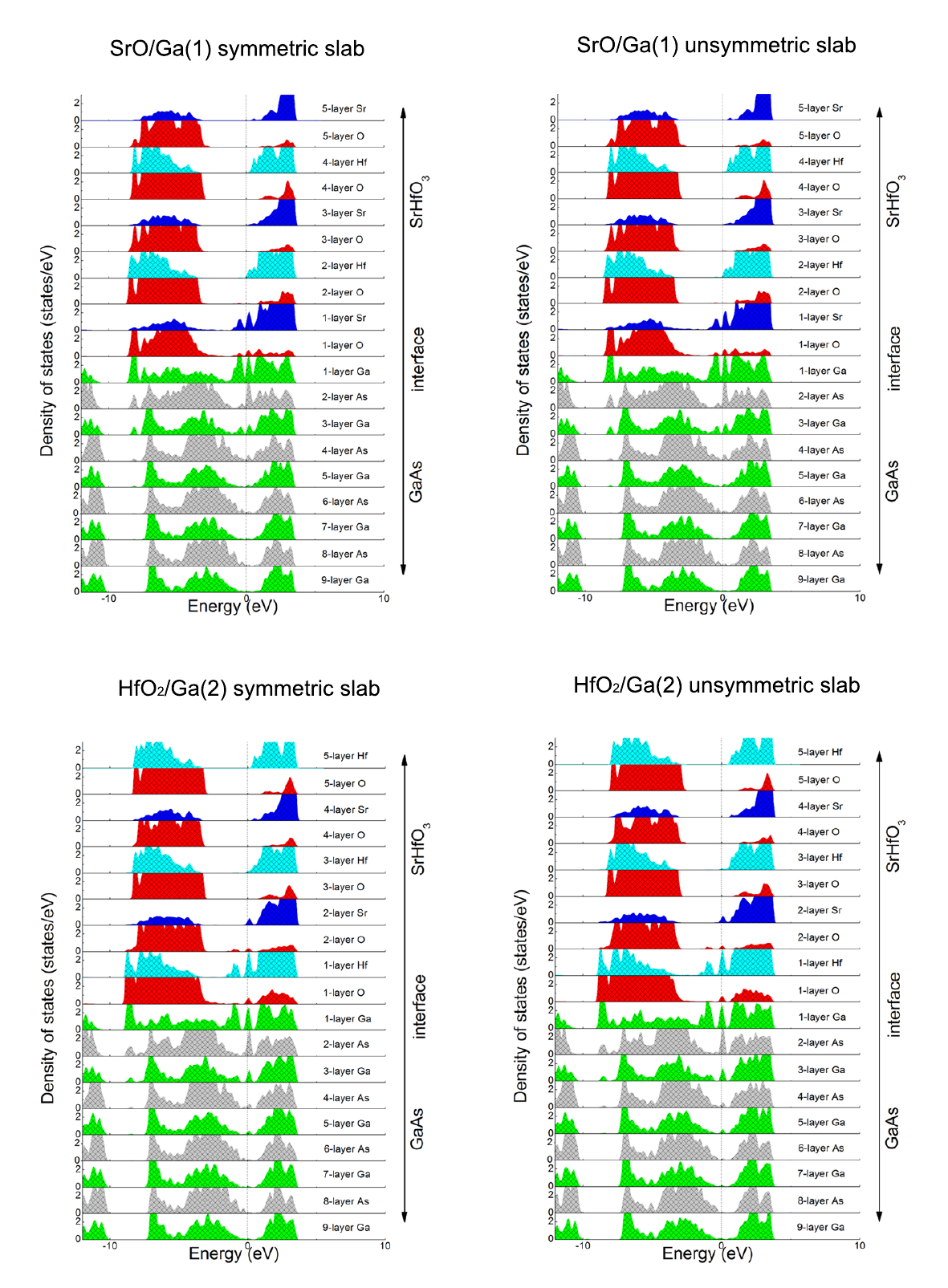}
\caption{Density of states of (001) $\rm SrHfO_3$/GaAs interfaces. The layer density of states for $\rm SrO/Ga(1)$ and $\rm HfO_2/Ga(2)$ configurations is presented. The vertical dotted line denotes Fermi level. The symmetric and unsymmetric slabs are considered.}
\label{figure6}
\end{figure*}

\begin{figure}
\centering
\includegraphics[width=8.5cm]{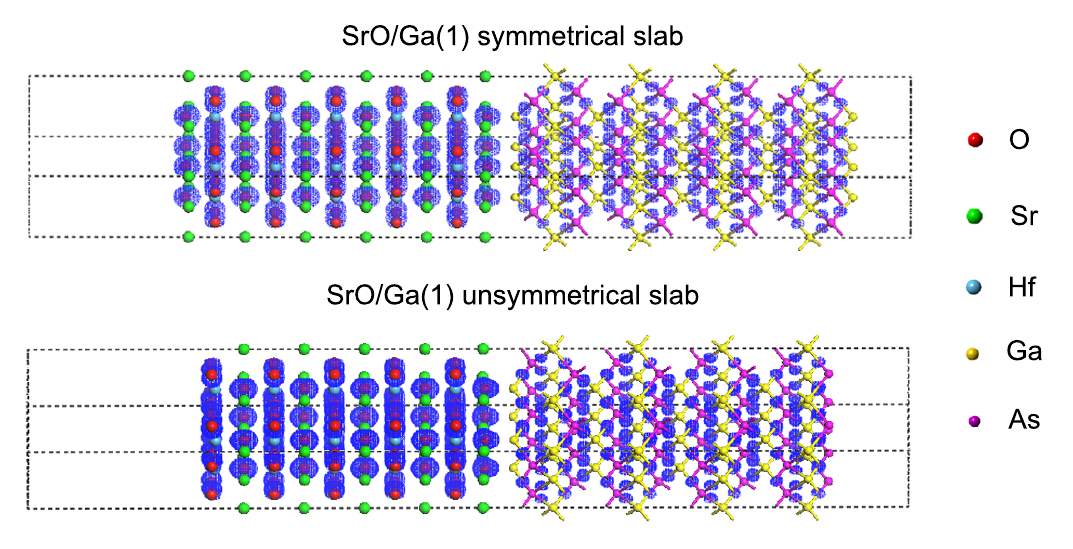}
\caption{The electron density difference for $\rm SrO/Ga(1)$ configuration with symmetrical and unsymmetrical slabs.}
\label{figure7}
\end{figure}
\begin{acknowledgements}
The work is supported by the Science Foundation from Education Department of Liaoning Province, China (Grant Nos. LF2017001 and LQ2017005).
\end{acknowledgements}
%

\end{document}